\documentclass[conference,compsoc]{IEEEtran}
\usepackage{fancyhdr}
\usepackage{graphicx}
\usepackage[bookmarks=true,hyperfigures=true,colorlinks=true,linkcolor=blue,citecolor=blue,pdfpagemode=fullscreen,plainpages=false,pdfpagelabels]{hyperref}
\usepackage[all]{hypcap}
\usepackage{cite}
\newcommand{\vbar}{\,|\,}
\title{Assurance 2.0: A Manifesto}
\author{\IEEEauthorblockN{Robin E. Bloomfield}
\IEEEauthorblockA{Adelard LLP and City, University of London\\London UK}\and
\IEEEauthorblockN{John Rushby}
\IEEEauthorblockA{Computer Science Laboratory\\
SRI International, Menlo Park CA USA
}}
\begin{document}
\maketitle
\pagestyle{plain}
\begin{abstract}

System assurance is confronted by significant challenges.  Some of
these are new, for example, autonomous systems with major functions
driven by machine learning and AI, and ultra-rapid system development,
while others are the familiar, persistent issues of the need for
efficient, effective and timely assurance.  Traditional assurance is
seen as a brake on innovation and often costly and time consuming.  We
therefore propose a modernized framework, ``Assurance 2.0,'' as an
enabler that supports innovation and continuous incremental assurance.
Perhaps unexpectedly, it does so by making assurance more rigorous,
with increased focus on the reasoning and evidence employed, and
explicit identification of defeaters and counterevidence.

\end{abstract}

\section{Introduction and Overview}

Assurance is often seen as a drag on innovation and as an activity
that is additional to (and generally comes after) the ``real work'' of
design and implementation.  We instead propose that assurance can be
an enabler for innovation and a constructive element in a holistic
design process.  However, if assurance is employed from the earliest
stages of design, it will necessarily be incomplete at those stages,
so we need some measures to indicate if we are headed in the ``right
direction'' and to help prioritize issues and solutions.
Counterintuitively, perhaps, we propose that the way to address these
and other concerns that we will introduce later, is by making
assurance more rigorous, in a framework that we call ``Assurance 2.0.''

This framework aims to support reasoning and communication about the
behavior and trustworthiness of engineered systems and, ultimately,
their certification.  It builds on the notion of an ``Assurance
Case,'' where claims about the system are justified by an argument
based on evidence.  In particular, it maintains a representation of
the \emph{structure} of the argument as a tree of claims linked by
argument steps and grounded on evidence (e.g., Figure \ref{defeat}) as
in \emph{Claims-Arguments-Evidence} (CAE) \cite{ASCAD} and \emph{Goal
Structuring Notation} (GSN) \cite{Wilson-etal96}\footnote{We use a
variant on CAE terminology: we say \emph{claim} where GSN says
\emph{goal}, we say \emph{argument step} where CAE says simply
\emph{argument} (and GSN says \emph{strategy}) and we use
\emph{argument} for the whole tree of claims and argument steps.  Our
diagrams use the CAE style.}, but strengthens it with increased focus
on the evidence and the reasoning (both logical and probabilistic)
employed, and on exploration and assessment of doubts and
``defeaters.''  We introduce the ideas in this section, using some
technical terms such as ``Confirmation Theory'' and ``Natural Language
Deductivism'' that are given in italics and are detailed (with
references) in later sections.

In current practice, steps in an assurance argument are often
\emph{inductive},\footnote{This is an unfortunate choice of words as
the same term is used with several other meanings in mathematics and
logic.} meaning the subclaims strongly support the parent claim, but
do not ensure it, as a \emph{deductive} step would.  In Assurance 2.0
we advocate that argument steps should be deductive, and this can
require additional evidence.  For example, argument steps often
iterate over some enumeration (e.g., over components, or over hazards)
and for this to be deductive we need evidence that the enumeration is
complete and that the claim distributes over its elements.  In cases
where it seems impossible to provide a deductive step, the ``gap'' in
reasoning must be acknowledged and given special attention.  To
support these recommendations, we advocate use of pre-analyzed
argument templates such as \emph{CAE Blocks}
\cite{Bloomfield&Netkachova14}, which provide mechanisms for managing
the \emph{side conditions} that are necessary to justify deductive
steps and excuse inductive ones.  This insistence that reasoning steps
should be ``as deductive as possible and inductive only as strictly
necessary'' is one of the ways in which Assurance 2.0 strengthens
traditional assurance; deductive reasoning steps ensure that doubts
have nowhere to hide and thereby help identify weak spots and focus
attention in productive directions.  Furthermore, this increased rigor
clarifies what must be accomplished at each step (indeed, there are
only five basic CAE Blocks: evidence incorporation, calculation,
decomposition, substitution, and concretion
\cite{Bloomfield&Netkachova14}) so that assurance developers are less
``bewildered by choice.''

Arguments are grounded on evidence and we advocate explicit assessment
of the ``weight'' of evidence offered in support of a claim.  It is
not enough for evidence to support a claim; it must also discriminate
between a claim and its negation or \emph{counterclaim} (weak evidence
could support either).  We recommend interpreting ``weight'' using
ideas and measures from \emph{Confirmation Theory}, which do exactly
this.  Again, this aspect of Assurance 2.0 is more demanding than
traditional estimates for the strength of evidence and requires
explicit consideration of counterclaims.  

Claims supported by sufficient weight of evidence may be used as
premises in a logical interpretation of the overall assurance argument
and when, in addition, all the reasoning steps are deductive, we have
a deductive thread from facts, established by evidence, to the top
level claim and thereby satisfy a benchmark for informal reasoning
known as \emph{Natural Language Deductivism} (NLD).

Although it is primarily motivated by practical considerations and by
experience with current methods, the Assurance 2.0 framework aligns
with modern developments in epistemology (notably, confirmation theory
and NLD), and we strengthen this alignment through use of
``indefeasibility'' as the criterion for justified belief (e.g., that
an assurance case establishes its claim).  For a belief to be
justified, the \emph{Indefeasibility Criterion} requires that we must
be sure that all identified doubts and objections have been
attended to and no credible doubts remain that would change the
conclusion.  We cannot be certain of our beliefs and there will always
be some residual doubts.  What makes the case indefeasible is that we
know about these residual doubts, have examined them, and made a
conscious decision about them.  Indefeasibility is lost when there may
be doubts that we do not know about, or doubts we do know about but
have not consciously addressed.

Doubts and objections are exemplified as \emph{defeaters}, so the
indefeasibility criterion applied to assurance cases requires a
comprehensive search for defeaters to the argument.  Once a potential
defeater has been identified, it must itself be defeated, meaning that
more detailed analysis shows that it is not, in fact, a defeater, or
that the system and/or its assurance case are adjusted to negate it.
In Assurance 2.0, we advocate that the search for defeaters, and their
own defeat, should be systematized and documented as essential parts
of the case (just as hazard analysis and the hazard log are essential
parts of safety engineering).  One systematic approach is through
construction and dialectical consideration of \emph{counterclaims} and
\emph{countercases}.  Counterclaims arise naturally in confirmation
measures and are discussed in Section \ref{evidential}, while a
countercase is an assurance case for the negation of the top claim and
is discussed in Section \ref{counter}.

Confirmation bias---the tendency to interpret information in a way
that confirms or strengthens our prior beliefs---is a natural concern
with assurance cases: after all, we are engaged in building a case to
support the system.  Competent and diligent external reviewers are
good defenses against confirmation bias, but are typically involved
only periodically and mostly toward the end of the development of a
case.  Several of the innovations in Assurance 2.0 are intended to
provide systematic mitigations against confirmation bias at every step
in the development of a case without the excessive conservatism that
leads to verbose cases with unnecessary evidence presented ``just in
case,'' and even to the rejection of good systems.  

The paper is organized as follows.  Section 2 describes the basic
structure of an assurance case argument and the criteria for
evaluating its soundness.  Section 3 discusses confidence in the case
and Section 4 provides brief conclusions.

\section{Arguments, Step by Step}

A key innovation in the development of modern assurance cases was the
idea of a ``structured safety case,'' introduced in the 1970s, that
required an \emph{argument} to explain how the design of the system
and the checks and tests performed during its development combine to
ensure safety.  Subsequent refinements in the 1990s led to the idea
that the argument itself should be structured, that is, organized
around goals or \emph{claims}, and grounded on \emph{evidence} about
the system.  Methods and notations such as GSN \cite{Wilson-etal96}
and CAE \cite{ASCAD} emerged at this time and support a body of
expertise and practice that thrives to this day.

The general structure of an assurance case argument is illustrated in
later diagrams, such as Figure \ref{defeat}.  An argument is organized
as a tree with two kinds of basic \emph{steps}: evidential (at the
leaves) and reasoning (interior), which are described in the following
subsections.  Mixed forms and cross links are also possible.

\subsection{Elementary Evidential Steps}

\label{evidential}

\begin{figure}[h]
\begin{center}
\includegraphics[width=1.8in]{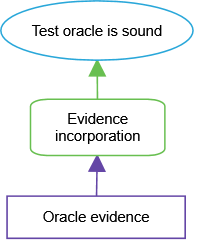}
\end{center}
\vspace*{-3ex}
 \caption{\label{soundoracle}Elementary Evidential Step}
\end{figure}

Let us begin with the most basic kind of argument step: one where some
item of evidence directly supports a claim.  To make things concrete,
we will suppose our examples are taken from a case in which random
tests are used to support a claim of reliability (certain cases for
nuclear systems are like this \cite{IAEA18:nuclear}).  One step in the
argument for this case will concern soundness of the test oracle: that
is, soundness of the means by which we judge the correctness of test
outcomes.  Figure \ref{soundoracle} portrays this step: at the top is
the (sub)claim that the oracle is sound (which will be backed by a
description of what it means for an oracle to be sound); at the bottom
is a description of the evidence for its soundness (which will be
backed by reference to files containing the actual evidence), and in
between is an argument that the evidence does indeed guarantee the
claim; we say that this argument is one for \emph{evidence
incorporation} and we refer to the whole argument step (i.e., claim,
argument, and evidence) as an \emph{evidential step}.

Implicit in the previous sentence is the idea that claims and
subclaims are logical propositions: that is, statements about the
``world'' (by which we mean the system of interest and its environment)
that may be true or false.  Evidence, on the other hand, is a
description or pointer to some observation or experiment in the world.
An argument for evidence incorporation documents a human assessment
that the evidence persuasively attests the truth of the claim.

This assessment may be informal, or it may employ some systematic
process.  In the latter case, it is usual to talk of \emph{weighing}
the evidence and of accepting the claim when the weight of its
supporting evidence crosses some threshold.  This raises the question
of how weighing is performed and what units are employed.  A standard
treatment uses probabilities: if $e$ is some evidence then $P(e)$ is
the probability of seeing this evidence.  This is
generally interpreted as a \emph{subjective probability}, that is, a
human judgment of likelihood expressed numerically from 0 (impossible)
to 1 (certain).  Similarly, $P(c)$ is the subjective probability that
the claim $c$ is true.  We might consider this the ``background'' or
\emph{prior} probability, which is then ``boosted'' by the evidence
$e$ to the \emph{posterior} probability $P(c \vbar e)$.  Thus, $P(c
\vbar e) > t$ for some $t$ might be considered a suitable criterion
for accepting $c$ on the basis of $e$.

Let us suppose that the evidence for soundness of our oracle is that
it was extensively validated against the previous version of the
system.  This seems like fairly strong evidence so we might make the
qualitative assessment that $P(c \vbar e)$ is ``high.''  However, a
critic might say that if the evidence is about a previous version of
the system, how relevant can it be to soundness of the oracle for this
version?  A sharp and general version of this question asks whether
the evidence can discriminate between a claim and its negation, or
counterclaim.  This suggests the weight of evidence should not be
based on $P(c \vbar e)$ alone, but should also consider the
\emph{difference} between this value and $P(\neg\, c \vbar e)$.
Difference can be measured as a ratio, or as arithmetic difference.

An attractive variant turns these conditional probabilities around:
instead of the posterior probability of the claim $P(c \vbar e)$, we
consider the \emph{likelihood} of the evidence given the claim, $P(e
\vbar c)$, and compare this to its likelihood given the counterclaim,
$P(e \vbar \neg\, c)$.  Likelihood and posterior probability are
related by Bayes' rule and so the choice of one over the other might
seem moot.  However, it is often easier to estimate the likelihood of
concrete observations, given a claim about the world, than vice-versa
(i.e., it is easier to estimate a likelihood than a posterior).
Furthermore, the likelihood $P(e \vbar c)$ has a more ``causal''
flavor---we think of (the property underlying) the claim causing the
evidence rather than vice-versa.

These ideas, and the general topics of evaluating and measuring
``weight of evidence,'' date back to the World War II codebreaking
work of Turing and Good \cite{Good:weight83}, where Good's
original measure for weight of evidence was $\log\frac{P(e \vbar
c)}{P(e \vbar \neg\, c)}$.  Today, these topics are studied in
Bayesian Confirmation Theory (a subfield of Bayesian Epistemology
\cite{Bovens&Hartmann03}) and many \emph{confirmation} (i.e., weight)
\emph{measures} have been proposed \cite{Tentori-etal07}.  Among
these, that of Kemeny and Oppenheim is popular: $$\frac{P(e \vbar c) -
P(e \vbar \neg\, c)}{P(e \vbar c) + P(e \vbar \neg\, c)}.$$ This
measure is positive for strong evidence, near zero for weak evidence,
and negative for counterevidence.

Returning to our example, we need to estimate the likelihood of the
evidence about the oracle (i.e., it exhibited good performance against
a previous version of the system), given a) the claim that the oracle
is sound, and b) the counterclaim that it is not.  An oracle evaluates
tests and their outcomes against requirements, so we need to ask
whether the requirements have changed between the previous and current
versions of the system.  Let us suppose the answer is ``yes, a
little.''  It's good that we asked, for the proffered evidence tells
us nothing about the performance of the oracle against those
requirements that have changed from the previous system (unless we
know more about the oracle structure and the modularity of the
requirements).  Without further evidence about the nature of the
requirements and the oracle, the Kemeny-Oppenheim measure is zero
and we conclude that the proffered evidence is of no value.

Although in most cases we do not advocate assessment of numerical
valuations for confirmation measures, nor their constituent
probabilities, we believe that informal consideration as was done here
(with ``qualitative'' assessments such as \emph{low}, \emph{medium},
and \emph{high}) can provide significant benefits in the evaluation of
evidence.

What are these benefits?  There are just a couple of ways in which a
well-formed
assurance case can be flawed or, as we say, \emph{defeated}
\cite{Goodenough-etal:ICSE2013}.  One is that the evidence supporting
a claim is inadequate to justify the confidence required; philosophers
call this \emph{undercutting defeat}.  It could be that the evidence
is merely insufficient (e.g., we did some testing, but not enough of
it) or it could be that its relationship to its claim has a gap or
flaw (e.g., the case just considered of an oracle evaluated against a
previous version of the system).  Confirmation measures, even when
assessed informally, provide rational quantification for the weight of
evidence and thereby guard against undercutting defeat.

The other kind of defeat is when there is evidence that contradicts a
claim; this is called a \emph{rebutting defeater}.  Confirmation
measures require consideration of the extent to which proffered
evidence supports counterclaims, and this should also invite
consideration of alternative evidence that could support the
counterclaim, or some other claim, and thereby guide a search for
rebutting defeaters within evidential steps.

Defeaters for an assurance case are rather like hazards for a critical
system, and just as the search for hazards is an essential element in
the engineering of critical systems, so the search for defeaters is an
essential element in the evaluation of assurance cases.  Confirmation
measures are an attractive tool in this search as they identify
both kinds of defeat in evidential steps and thereby provide a
valuable and necessary antidote to \emph{confirmation bias}, which
some consider an endemic vulnerability in assurance cases
\cite{Leveson11:JSS}.  

This section considered only elementary evidential steps; a less
elementary step may incorporate several items of evidence in support
of a single claim.  The overall confidence measure can then involve
conditional probabilities and likelihoods for evidential items that
are not independent of each other.  Tools for Bayesian Belief Nets
(BBNs) can assist in construction and evaluation of numeric models for
these circumstances.  Although we do not advocate numerical
assessments for the probabilities involved, ``what if'' experiments
with a range of possibilities can prove very enlightening.  An example
is given in \cite{Rushby:AAA15}.

\subsection{Elementary Reasoning Steps}

\begin{figure}[h]
\begin{center}
\includegraphics[width=3.0in]{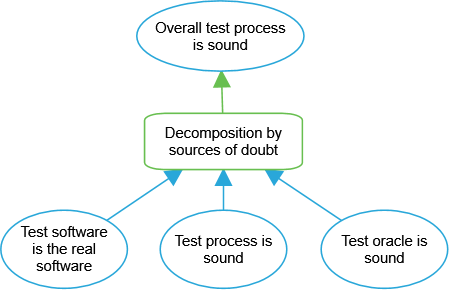}
\end{center}
\vspace*{-3ex}
 \caption{\label{reasons}Elementary Reasoning Step}
\end{figure}

We have considered an elementary evidential argument step---one where
we assess the extent to which evidence supports a claim---and now turn
to a similarly elementary reasoning step---one where several
(sub)claims combine to support a parent claim.  Figure \ref{reasons}
illustrates such a step.  Here we suppose we have three subclaims
concerning a test procedure, each supported by evidence or an entire
subargument (these are not shown): one asserts that the test oracle is
sound (as in the previous section), another that the test procedure is
sound, and the third that the tested software is the actual software.
The step asserts that if these three subclaims are true, then we may
conclude that the overall test process is sound.  Each subclaim will
be backed by a description of what it means and collectively they will
be bound together by an argument that they ``lead to'' the parent
claim.

We say the subclaims ``lead to'' the parent because we have not yet
established the relationship that is intended.  In some early
interpretations for an assurance case, the intended relationship was
structural rather than logical: it simply indicated that the case for
the parent claim decomposed into subcases for each of the 
subclaims.  In modern interpretations, the intended relationship is
logical but it may be deductive (i.e., the subclaims imply or entail
the parent claim) or inductive (i.e., the subclaims ``suggest'' the
parent claim).  In diagrammatic presentations, an annotation on the
central argument box can indicate which of these is intended.

\begin{figure}[h]
\begin{center}
\includegraphics[width=3.0in]{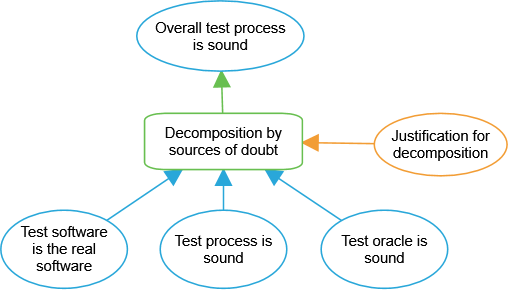}
\end{center}
\vspace*{-3ex}
 \caption{\label{reasons-just}Reasoning Step with Side Condition}
\end{figure}

When a deductive interpretation is indicated, the argument must make
the case that the subclaims truly entail the parent claim.  Sometimes
a convincing case can be made with no additional information, but
often an additional subclaim will be needed to substantiate the case.
Logically, this additional subclaim is just like the others and
conjoins with them to entail the parent claim; however, it is
contextually somewhat different, so we call it a ``side condition'' or
``side claim'' (or sometimes an ``assumption'') and draw it in a
different position and color (but same shape), as shown in Figure
\ref{reasons-just}.  In this case, we are claiming the three
conditions considered in the original subclaims are the only threats
to overall soundness of the testing process and the side condition,
which asserts this, will need to be supported by evidence, akin to
hazard analysis, to justify it.

Observe that what we are doing here is constructing a theory for sound
testing, but we are doing it in an \emph{ad hoc} manner during
construction of a larger assurance case.  In current practice,
assurance cases often contain subcases that develop \emph{ad hoc}
theories or models in this way.  For Assurance 2.0 we recommend that
such theories and models should be developed explicitly and separately
from the overall case.  This allows them to be suitably reviewed and
validated, and reused in other cases.  For example, we could have an
explicit theory for sound testing that would develop and justify all
the hazards to a test campaign.  Figure \ref{reasons-just} might still
look the same, but its subclaims (and there would probably be several
more of these) would be supplied by decomposition on the theory for
sound testing and the side claim and narrative justification would
reference that theory.


In Assurance 2.0, we advocate that all reasoning steps eventually
should be deductive as this raises the bar on the quality of
argumentation required and is necessary to satisfy the
indefeasibility criterion for justified belief in the overall
argument.  Dually, challenges to deductiveness can provide a
systematic basis to the search for defeaters, whose energetic pursuit
is a rational guard against hubris and confirmation bias in the
construction of assurance cases.  The challenge and response to
defeaters of this kind often goes hand-in-hand with (re)formulation of
side conditions: strengthening a side condition, and its supporting
evidence, is one way to respond to a successful defeater.

Observe that some elements that may appear in claims (e.g., an
expression like $\frac{x}{y}$) may not even ``make sense'' unless a
suitable ``assumption'' side claim is true (e.g., $y \neq 0$).  Such a
claim may be established in one place but used in many others, so the
overall argument is no longer  a tree.

The precision and rigor we advocate via deductiveness needs to be
reconciled with the need for concise communication and constructive
progress during system development.  Thus, we accept that the case
will be incomplete and argument steps may be inductive during the
early stages of system development and assurance exploration.  But it
is desirable that tools should assist in keeping track of these
transitional compromises.  A technique for tools based on deductivism
would be to supply inductive steps with a nugatory ``something missing
here'' side claim that is asserted to make the step deductive, but is
unsupported by evidence.  This allows progress, while the unsupported
side claim acts as a constant reminder of imperfection in this
argument step.

This discussion has considered only elementary reasoning steps: those
where a claim is supported by subclaims.  In less elementary reasoning
steps, a claim may be supported by a combination of subclaims and
evidence.  The most useful construction of this kind is best
interpreted not as a reasoning step, but as an evidential step with
side claims that function as assumptions.  Reference
\cite{Rushby:Shonan16} provides more discussion of these topics.

\section{Soundness and Confidence Assessment}

In Assurance 2.0, the interpretation that we apply to an assurance
case is a systematic instance of ``Natural Language Deductivism''
(NLD) \cite{Groarke99}, which regards its informal argument as an
approximation to a deductively valid proof.  NLD differs from proof in
formal mathematics and logic in that its premises are ``reasonable or
plausible'' rather than certain, and hence its conclusions are
likewise reasonable or plausible rather than certain.  Our
requirements that evidential steps cross some threshold for
credibility (as assessed by a confirmation measure, for example), that all
reasoning steps are deductive, and that a thorough search for
defeaters persuades stakeholders that the case is indefeasible,
systematizes what it means for the premises to be ``reasonable or
plausible'' and thereby give us confidence that the overall argument
is sound and the top claim is true.  But then we might ask, how much
confidence does it give us, and how much do we need?

Some assurance cases may be more persuasive than others, and not all
(sub)systems need the highest levels of assurance: indeed, several
standards speak of ``Safety Integrity Levels'' (SILs) from 1 (low) to
4 (high) \cite{IEC61508} or ``Design Assurance Levels'' (DALs) from E
(low) to A (high) \cite{DO178C}.  Thus, we need ways to assess
confidence in a case, and principled ways to organize cases so that
the lower SILs and DALs are easier and cheaper to achieve.  The
confidence we need depends on the nature of the claim and the decision
being made.  In some cases (we call them ``explicit''), the claim may
include a numeric estimate for some parameter (e.g., reliability) and
our confidence then reflects epistemic uncertainty in this quantity.
In others (we call them ``implicit''), the claim may be absolute and
confidence is then a separate estimate of its likelihood expressed as
a subjective probability.  For example, the claim may be that the
system has no faults, and confidence in this claim (sometimes called
``probability of perfection'' \cite{Littlewood&Rushby:TSE12}) can be
used to estimate long run survival without critical failures by means
of Conservative Bayesian Inference (CBI) \cite{Zhao-etal19:AVtests}.

\begin{figure}[ht]
\begin{center}
\includegraphics[width=\linewidth]{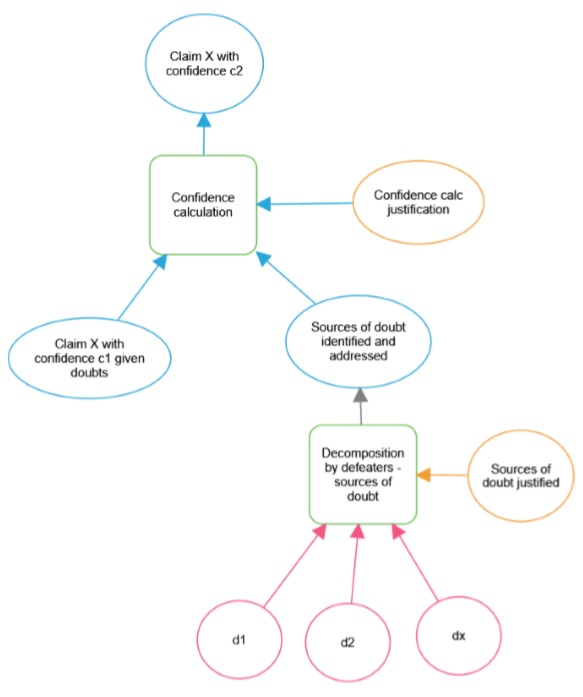}
\end{center}
 \caption{\label{confbuildpat}Confidence Building Pattern}
\end{figure}

Confidence should be related to the number and severity of potential
defeaters that have been considered and eliminated or mitigated.
Identification and treatment of defeaters are often addressed
implicitly as safety engineers question whether the arguments they put
forward are valid and they adjust and correct the case as necessary.
In Assurance 2.0, we attempt to record these defeaters and their
treatment as part of the assurance csse and we use these records in
assessing confidence.  However, some defeaters are simply errors or
deficiencies in reasoning (e.g., well-known fallacies, such as
reasoning from the specific to the general, or the wrong instantiation
of an applicable rule) \cite{Greenwell-etal06} and in Assurance 2.0 we
attempt to eliminate these ``by construction'' through the use of
predefined CAE Blocks and do not record them.  Others are due to
oversights or incorrect reasoning about behaviors and dependencies and
we attempt to detect and record those during construction by explicit
consideration of sources of doubt (i.e., undercutting defeaters).  A
pattern of reasoning (i.e., a macro on blocks) that we have introduced
to do this is shown in Figure \ref{confbuildpat}.

Here, the claim $X$ initially (at lower left) has confidence $c_1$
associated with it, but this is contingent on doubts and assumptions
concerning the claim.  The right-hand subcase assesses these and the
``confidence calculation'' block evaluates a revised confidence $c_2$
for claim $X$.  The side claim has to justify why the theory or
``calculus'' of confidence used is valid and how it supports the top
claim.  Meanwhile, the doubts and missing assumptions and other
potential defeaters contributing to this calculation are examined and
addressed in the decomposition block at the lower right of the case,
and the sum of their significance is supplied to the subclaim of the
confidence calculation.  The side condition to this block must justify
that those identified are all and only the relevant sources of doubt.
Notice here that the subcases $d_1, d_2,\ldots,d_x$ at the bottom of
this case concern sources of doubt that have not yet been addressed or
mitigated (i.e., defeaters) such
as ``test oracle is unsound'').  This is indicated by the red color of
the claims and arrows, whose semantics are that any one of them
defeats the argument above.  As development of the case continues, the
defeaters will be eliminated or mitigated and their claims will become
``positive'' and the red color replaced by blue.  The human factors
and dynamics of how defeaters should be presented and archived is a
topic of investigation.  Takai and Kido describe the methods used in
\emph{Astah GSN} \cite{Takai&Kido14}.

There are several ways to use the pattern of Figure
\ref{confbuildpat}.
\begin{enumerate}
\item Claim $X$ could be an absolute proposition about functional
behavior (e.g., there are no deadlocks) or a probabilistic one that
includes aleatory uncertainty (e.g., 95\% confidence that the
reliability is better than three nines).\footnote{Aleatoric (or
aleatory) uncertainty is uncertainty \emph{in} the world: if I toss a
fair coin 100 times, the number of heads is subject to aleatoric
uncertainty; epistemic uncertainty is uncertainty \emph{about} the
world: if I give you a coin and invite you to toss it 100 times, there
is additional uncertainty about the number of heads because you do not
know if the coin is fair or not.}  Assumption and other epistemic
doubts are then explored by identifying defeaters.  If the subcase to
the right shows that these are all mitigated and their impact is below
some threshold then confidence in $X$ remains the same and $c_2 =
c_1$.  This form is the most common when we deal with epistemic doubts
qualitatively.

\item If evaluation of the defeaters reveals significant impact, then
$c_2$ will be larger than $c_1$.  This calculation might be done
quantitatively in terms of some descriptive scale (e.g., \emph{low},
\emph{medium}, and \emph{high}).

\item We may have probability distributions for $c_1$ that can be
combined in a conservative Bayesian manner with the doubts.  This
could be done either qualitatively as in the example leading up to
Figure \ref{defeat} or quantitatively as in the example of Figure
\ref{bbn2}.

\end{enumerate}


We next consider how assessments of confidence for individual claims
can be accumulated and propagated through a case to yield an
assessment of confidence in the top claim.

\subsection{Issues in Assessing Confidence}
\label{qual}

A natural measure for confidence in the claim of an evidential
step is $P(c \vbar e)$; as explained in Section \ref{evidential}, we
do not use this as a measure for the \emph{weight} of evidence because
that must also account for the ability of the evidence to discriminate
between the claim and a counterclaim, but once the evidence has been
accepted on the basis of its weight, it is reasonable to use $P(c
\vbar e)$ as our confidence in its claim.

Next, we need a method to ``combine'' the confidence measures from the
evidentially supported subclaims of a reasoning step to yield a
confidence measure for its parent claim, and so on up to the root of the
tree where we obtain a confidence measure for the top claim.
Probability and logic build on completely different foundations and
their combination is difficult.  Graydon and Holloway
\cite{Graydon&Holloway:quant17} examined 12 proposals for using
probabilistic methods to quantify confidence in assurance case
arguments: 5 based on Bayesian Belief Networks (BBNs), 5 based on
Dempster-Shafer \cite{Shafer76} or similar forms of evidential
reasoning, and 2 using other methods.  By perturbing the original
authors' own examples, they showed that all the proposed methods can
deliver implausible results.

However, in Assurance 2.0 we have a very simple special case.
Ideally, all our reasoning steps are deductive conjunctive
implications (i.e., definite clauses), so confidence in a parent claim
is given by the product of confidence in the subclaims (provided they
are independent)\footnote{If the subclaims are not independent, then
we will need to use an approach such as BBNs to assess their
combination.  Note that our use of BBNs is simpler than those examined
by Graydon and Holloway because we apply BBNs only to individual
reasoning steps (this is because we assess soundness separately from
confidcence).}.  Iterating this over the whole argument tree,
confidence in the top claim is the product of confidence in all the
evidentially supported claims.  If we have reasoning steps that are
not deductive, then it is sound (though often highly conservative) to
calculate \emph{doubt} (i.e., $1 - \mbox{confidence}$) in a parent
claim as no worse than the sum of doubts of its subclaims
\cite{Adams98}.

Confidence in individual claims may itself be expressed qualitatively
(e.g., \emph{low}, \emph{medium}, and \emph{high}) and so it will be
necessary to develop plausible rules for the ``product'' of such
estimates (e.g., the product of 15 to 25 \emph{high}s yields
\emph{medium}).  Adjusting a case, or a case template, for different
SILs can be accomplished by weakening claims, and by reducing the
quantity or quality of evidence demanded; this may in turn allow some
subclaims and their supporting argument to be eliminated: e.g., if we
replace static analysis by human inspection, we no longer need a
subcase for soundness of the static analyzer (but we will need a
subcase for reviewer efficacy).  Subclaims should not otherwise be
removed, for that necessarily makes the case inductive, but we could
reduce the threshold at which minor caveats and defeaters are
considered mitigated.

In the early stages of system development, the assurance case may be
very incomplete yet we would still like to get guidance on areas where
attention should be focused.  One possibility is to assign exaggerated
numerical assessments for projected confidence in various subclaims
(e.g., 95\% for \emph{high} estimated confidence, 5\% for \emph{low},
and 1\% for a nugatory side claim) and then ``run the numbers'' and do
``what if'' exercises to learn where the largest impacts reside.  The
tools supporting these calculations could also take challenges and
defeaters into account: a subclaim that has not been challenged would
have its confidence reduced, and undefeated defeaters would do the
same.  
Note that confidence is a ``weakest link'' phenomenon.  Thus, we do
best when we have approximately the same high level of confidence in
each claim.


\subsection{Explicit Representation of Confidence}
\label{quant}

Next, we consider an example where confidence is an explicit part of
the top claim and show how we use aspects of CAE Blocks and the
confidence pattern of Figure \ref{confbuildpat} to develop a case.  In
addition to confidence, this example also illustrates a more complex
development, where defeaters and counterclaims play an important part.

The example is a case based on statistical testing, as used in certain
nuclear applications.  The idea is that random tests that follow the
``operational profile'' can justify a reliability claim, such as
probability of failure on demand (\emph{pfd}) \cite{IAEA18:nuclear}.
It is also similar to many examples in autonomous systems where we
wish to build on simulation and field trials to support, or not, a
claim of safety related reliability.

In the example, we
\begin{itemize}
\item First focus on evidence integration and claim definition, then

\item Apply the confidence pattern to deal with defeaters,

\item Extend the case to reason quantitatively about epistemic
confidence, and

\item Apply again the confidence pattern as the quantitative reasoning
brings new defeaters.
\end{itemize}

\subsubsection{Evidence integration and claim definition}

\begin{figure}[h]
\begin{center}
\includegraphics[width=1.7in]{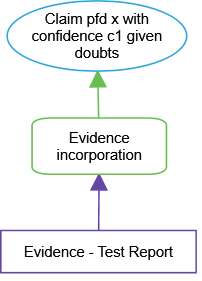}
\end{center}
\vspace*{-2ex}
 \caption{\label{init}Initial Case for Statistical Testing, with Doubt Annotation}
\vspace*{-1ex}
\end{figure}

In this example the initial evidence offered is a report of the tests
performed.  The analyst reviews this and integrates it into an
assurance case justifying a claim for a certain \emph{pfd} $x$ that is
held with confidence $c_1$, based on what the analyst considers to be
a well-known theory of statistical testing (e.g., \cite[Annex
I]{IAEA18:nuclear}).  This initial assurance case is shown in Figure
\ref{init}, where the top claim is a predicate that could be used in a
larger case, such as one that combines reasoning about reliability
with evidence for correctness.

However, on reflection, or under challenge, the analyst decides that
this initial case is subject to significant doubts because the
argument for evidence incorporation does not reach the threshold for
indefeasibility: for example, can we be sure the assumptions
underpinning the theory of statistical testing are satisfied?  In a
tool-supported environment for Assurance 2.0, there would be ways to
indicate this potential defeater but, for the diagrammatic
representation used here, the analyst simply notates the claim as one
with given doubts.

\begin{figure}[h]
\begin{center}
\includegraphics[width=2.4in]{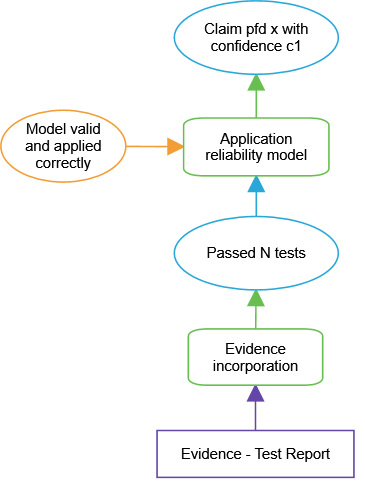}
\end{center}
 \caption{\label{sepn}Separation of Facts from Test Report and
 Inference of Reliability}
\end{figure}

The root problem is that the evidence incorporation step combines both
the extraction of facts from the test report, and their analysis and
interpretation with respect to a model of statistical testing.
Consequently, in Figure \ref{sepn} these two aspects are separated:
evidence incorporation extracts the purported facts, namely that the
system passed $N$ tests, and a substitution block provides the
argument that these justify the top claim, with a side claim (that
will eventually need to be justified by its own evidence) to support
the validity and correct application of the underlying model for
statistical testing and reliability.

\begin{figure}[t]
\begin{center}
\includegraphics[width=2.3in]{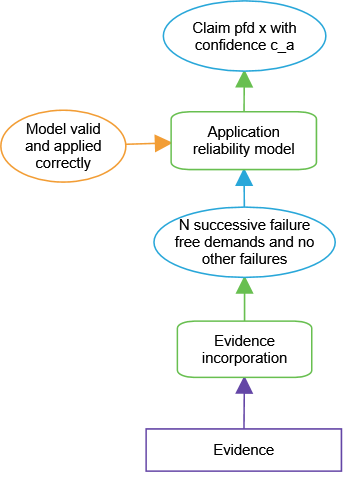}
\end{center}
\vspace*{-3ex}
 \caption{\label{precise}More Precise Claims}
\end{figure}

In weighing the test evidence, confirmation measures will invite us to
consider whether ``Passed $N$ tests'' could be true of an unreliable
system as well as a reliable one.  We realize that if we don't mind
failures along the way, then more or less any system will eventually
pass $N$ tests.  Thus, we see the need for a more precise
interpretation: namely, that the tests demonstrated $N$ failure-free
demands in succession, and that no other failures were observed.
Consequently, this claim should be changed to ``$N$ successive
failure-free demands and no other failures.''  If the evidence can
support this claim (as opposed to a weaker claim where some failures
may have been observed) then we can retain Figure \ref{sepn} as our
assurance case, but with the claim ``Passed $N$ tests'' replaced by
the more precise form, as shown in Figure \ref{precise}.

Consideration of the side claim concerning validity of the reliability
model and its application forces realization that confidence in the
top claim is with respect to \emph{aleatoric uncertainty} (based on
the extent of testing) and this is reflected in the revised top claim
where $c_a$ replaces $c_1$.\footnote{$c_1$ and $c_a$ are numbers, but
they are annotated with descriptions of their interpretation and it is
these that change.}

Consideration of indefeasibility, confirmation measures, and side
claims suggested improvements in the case; we now look at defeaters.

\subsubsection{Applying the confidence building pattern}

Further reflection, or challenging peer review, might ask how do we
know that the tests were performed correctly, and that issues such as
correctness of the test oracle were addressed appropriately?  The
analyst recognizes that these are legitimate defeaters and the case
needs to be strengthened by using the theory and pattern for test
performance previously illustrated in Figure \ref{confbuildpat}.  This
leads to a new assurance case (not shown) in which Figure
\ref{precise} is a subcase dealing with reliability and confidence,
and an elaborated version of Figure \ref{confbuildpat} is a subcase
dealing with soundness of the overall test procedure.

\begin{figure}[ht]
\vspace*{-1ex}
\begin{center}
\includegraphics[width=\linewidth]{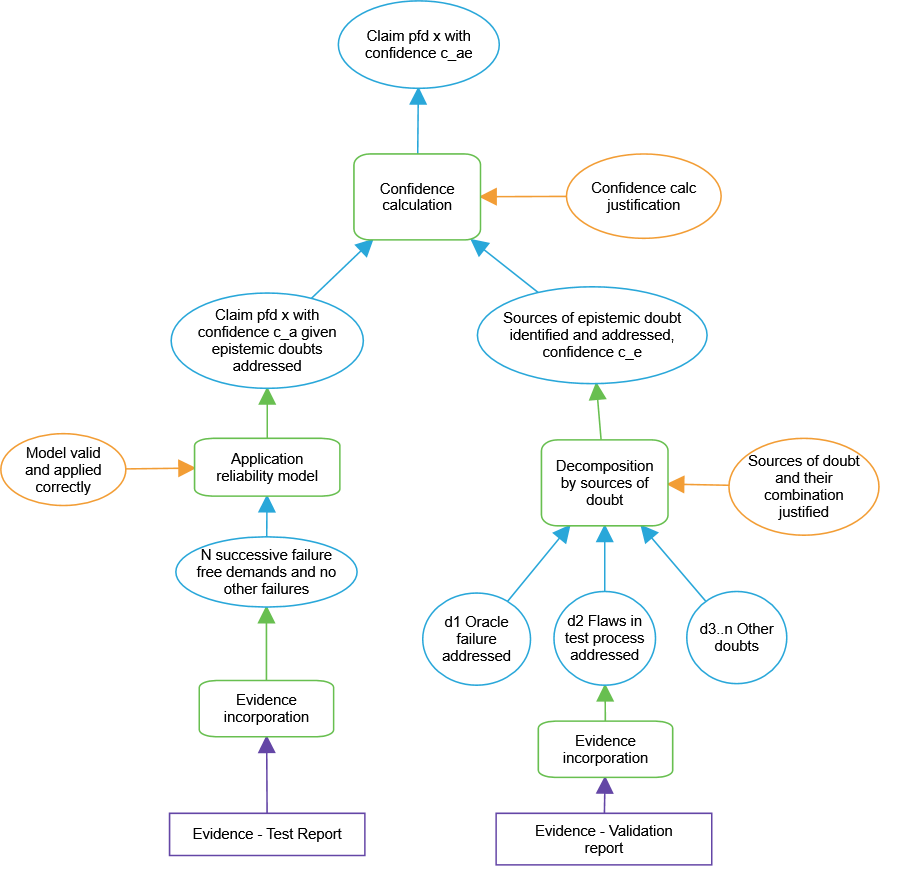}
\end{center}
\vspace*{-2ex}
 \caption{\label{defeat}Showing Defeaters Have Been Incorporated}
\end{figure}

A slight variant, which is appropriate because the top claim
explicitly states the confidence associated with the \emph{pfd} $x$,
is to interpret Figure \ref{precise} as a subcase dealing with
\emph{aleatoric} uncertainty and an elaborated Figure
\ref{confbuildpat} as a subcase dealing with \emph{epistemic}
uncertainties.  This approach is shown in Figure \ref{defeat} in which
the defeaters of Figure \ref{confbuildpat} have been addressed and
become positive claims.  Note that this and subsequent examples are
not complete cases: some claims lack supporting evidence.

Figure \ref{defeat} provides a pattern in which we separate reasoning
about aleatoric doubts (left-hand leg) from that about epistemic
doubts (right-hand leg).  However, we may sometimes need to reason
about aleatoric and epistemic aspects within the same framework, as
when we wish to model their interactions and dependencies.  This is
illustrated in the next example.

\subsubsection{Extending the confidence reasoning to quantified models
of epistemic doubts}

\begin{figure}[ht]
\vspace*{-1ex}
\begin{center}
\includegraphics[width=\linewidth]{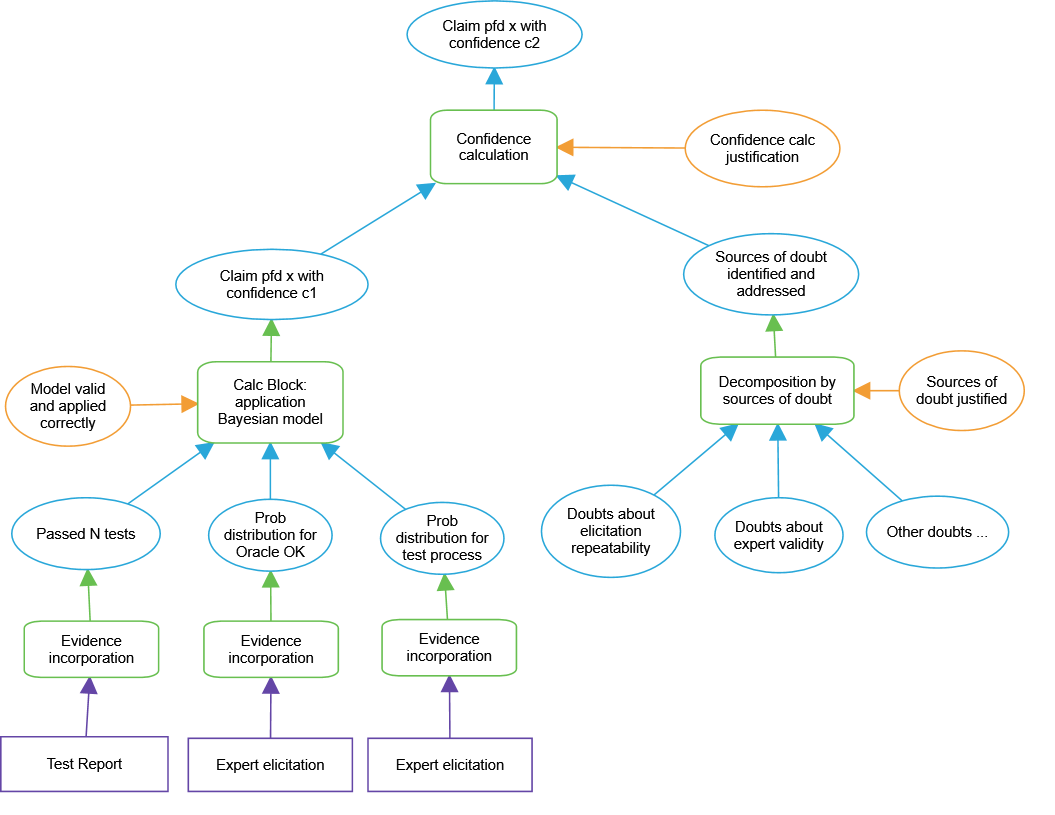}
\end{center}
\vspace*{-2ex}
 \caption{\label{bbn2}Incorporating BBN Modeling}
\end{figure}

If we were able to provide a quantified judgment of our confidence in
the soundness of the oracle and the test process in the form of
conditional probability distributions, then we could combine them in a
BBN model, as illustrated by the left-hand leg in Figure \ref{bbn2}.

Here, the leg uses Bayesian reasoning to provide a probability
distribution for the property of interest, and from that derives a
confidence figure in the claimed \emph{pfd} $x$.  There is a new side
claim that requires justification for the application and validity of
the BBN model.

\subsubsection{Seeking defeaters in the extended pattern}

We again apply the pattern of Figure \ref{confbuildpat} as using the
BBN model approach brings its own defeaters.  The right-hand leg of
Figure \ref{bbn2} applies the pattern to deal with defeaters to the
BBN approach; it has identified two concerning elicitation of the
probability distributions from experts: namely, its validity and
repeatability.  Furthermore, we have a side claim asserting these are
the \emph{only} sources of doubt.  We are not sure they are, so this
argument step is inductive; we choose to represent this by adding a
nugatory third ``other doubts'' claim.

These defeaters are formidable: it is seldom credible that we can
derive full conditional distributions as needed here.  If we can, then
the benefit is that the confidence calculated by the left-hand leg may
be much greater than can be supported by weaker assumptions and
conservative calculations as in Section \ref{qual}.  An intermediate
position in the tradeoff between confidence in the claim and doubts
about assumptions is to reduce criticality of the claim and increase
confidence in its reduced form: if we are 90\% confident that a
subcase establishes SIL3 then, with some additional modeling and
assumptions, we might become 99\% confident that it is better than
SIL2, and this could be sufficient to argue that it meets the
evidential threshold.  If challenged to deal with the remaining doubt,
we could use a chain of confidence \cite{Bishop-etal:TSE11} that
combines a firm judgment about the 99\% with a conservative judgment
about the other 1\%.

\subsubsection{Countercases}
\label{counter}

Another way of identifying defeaters or sources of doubt is to develop
an explicit countercase that aims to refute the claim under
consideration.  This task could be assigned to a different team,
which, given its different viewpoint, might generate challenging and
unexpected defeaters for the base case.  There is some tension here: a
totally independent countercase might have an argument structure
completely different to the base case, and thereby generate irrelevant
defeaters.  However, there seems to be a useful initial transformation
from a case to a parallel countercase (and vice-versa): mitigated
defeaters become claims and claims become a source of defeaters.  Even
so, a countercase would be very different from a positive case as only
one strong defeater would be needed to substantiate the counter
claim. This is similar to reasoning about security: the attacker only
needs to find one vulnerability, whereas the defender needs to defend
against all.



Instead of reasoning about the negation of the claim, it may be more useful
to argue several different diverse but related top-level claims
(e.g., different risk-based claims, claims of critical defect-freeness,
operation in a different environment).  As with explicit defeaters, the
use and utility of counter cases or diverse cases is one of the novel
areas that will be evaluated as Assurance 2.0 becomes deployed.

\section{Conclusion}

We have described and illustrated Assurance 2.0, whose purpose is to
respond to the assurance challenges posed by recent developments in
system design and deployment, and to provide a framework in which
assurance can become more dynamic and can enable innovation and greater
automation.  Assurance 2.0 retains the argument structure of Assurance
Cases and can build on much recent and current research and tooling
for these.  Where it differs is in stressing rigor in assessment of
the evidence and reasoning employed, and a focus on combatting
confirmation bias through active exploration and recording of
potential defeaters.

Key elements of Assurance 2.0 are:
\begin{itemize}
\item Use of a limited repertoire of five building ``blocks'' for
arguments, with an associated framework that reduces the bewildering
choice of free-form arguments by guiding assurance case development
toward productive directions, and that eliminates many errors ``by
construction'' through use of rigorous side conditions.

\item Cleanly separating the development and description of
``models,'' such as for the system environment, and ``theories'' such
as for the sound and productive use of static analysis, from the
assurance case itself.  This allows the case to focus on assembly and
evaluation of claims, arguments and evidence concerning these externally
described artifacts.

\item Use of an Indefeasibility Criterion for justified belief that
frames the notion of defeaters, both undercutting and rebutting, and
motivates construction of arguments that are predominately deductive,
an approach known as ``Natural Language Deductivism'' (NLD).

\item Use of Confirmation Measures to evaluate the strength of
evidence and arguments.  It is not enough for evidence to support a
claim; it must also discriminate between a claim and alternatives,
including its negation or counterclaim.

\item An approach to reduce confirmation bias through active search
for defeaters and a methodology for doing so by means of counterclaims
and countercases.

\end{itemize}

Current assurance cases have served traditional safety-critical
systems well \cite{Rinehart-etal:NASA15}, but we have observed them
floundering when confronted by radically new challenges such as
autonomous systems driven by machine learning, by new stakeholders
such as the AI community, and by applications with a security focus.
Assurance 2.0 renews the original focus of assurance by asking for a
natural language explanation why the proposed system satisfies the
properties claimed for it, while reinforcing this with the rigor of
NLD, and challenging it through systematic search for defeaters.  A
completed Assurance 2.0 Case attests to the relevance and strength of
its evidence and the deductive validity of its reasoning (although
inductive steps may be used if absolutely necessary), and also records
the defeaters to which it has responded, thereby establishing not
merely its plausibility but its soundness and indefeasibility.

We believe we are not alone in observing and diagnosing weaknesses in
current approaches to assurance, nor in our prescriptions for
improvement.  For example, Viger \emph{et al} favor deductivism
\cite{Viger-etal:Safecomp20} and Denney and Pai
\cite{Denney&Pai:safecomp13} and also C\^{a}rlan and Ratiu employ
formalised templates \cite{Carlan&Ratiu:Safecomp20}.

We have used ideas underlying Assurance 2.0 with some success in
training several groups of engineers and managers and applied them in
research projects with regulators and industry.  

For the future, we hope to see application of these ideas to
significant modern systems, supported by training across a wide range
of disciplines and the development of constructive tool support.  The
formal nature of the reasoning and evidential analysis that underlies
Assurance 2.0 should enable productive interaction with tools for
logical and probabilistic reasoning and formal argumentation, together
with novel automation in the search for defeaters, the construction of
cases and countercases, and the management and representation of
dialectical examination.  We plan to prototype and evaluate the
approach (e.g., following ideas of Graydon \cite{Graydon:EDCC20}) in
industrial applications and research projects including the DARPA
ARCOS program.  As we develop more material and experience with
Assurance 2.0 we will publish this on
\url{claimsargumentevidence.org}.

\subsubsection*{Acknowledgments} We thank colleagues at Adelard LLP,
City, University of London, and SRI for many stimulating discussions
on these topics.  This work was funded by Adelard and by SRI.

\vspace*{-0.8ex}



\end{document}